\documentstyle[12pt]{article}
\let\tilde=\widetilde
\let\hat=\widehat

\newcommand\commentout[1]{}
\newfont{\bbd}{msbm10 scaled\magstep1}
\def\Complex{\hbox{\bbd C}}
\def\Integer{\hbox{\bbd Z}}
\def\Proj{\hbox{\bbd P}}
\def\Real{\hbox{\bbd R}}
\def\calA{{\cal A}}
\def\calB{{\cal B}}
\def\calC{{\cal C}}

\def\calP{{\cal P}}

\newcommand\utilde{{\tilde u}}
\def\s{\sigma}

\def\tensor{\otimes}
\def\div{\mathop{\rm div}\nolimits}
\def\End{\mathop{\rm End}\nolimits}
\def\mult{\mathop{\rm mult}\nolimits}
\def\sov{\mathop{K}\nolimits}
\def\trace{\mathop{\rm Tr}\nolimits}
\def\half{\frac{1}{2}}
\def\texthalf{{\textstyle\frac{1}{2}}}

\def\tr{\mathop{\hbox{\rm tr}}\nolimits}

\def\one#1{#1^{\raise7pt\hbox{$\scriptstyle\!\!\!\!1$}}\,{}}
\def\two#1{#1^{\raise7pt\hbox{$\scriptstyle\!\!\!\!2$}}\,{}}
\def\three#1{#1^{\raise7pt\hbox{$\scriptstyle\!\!\!\!3$}}\,{}}
\def\phi{\varphi}

\def\a{\alpha}
\def\b{\beta}
\def\g{\gamma}

\def\s{\sigma}

\def\comment#1{}

\def\id{\hbox{{1}\kern-.25em\hbox{\rm l}}}
\newtheorem{thm}{Theorem}[section]
\newtheorem{lem}[thm]{Lemma}
\newtheorem{prop}[thm]{Proposition}

\newcommand\thmref[1]{Theorem~\ref{#1}}
\newcommand\propref[1]{Proposition~\ref{#1}}
\newcommand\secref[1]{Section~\ref{#1}}

\newcommand\lemref[1]{Lemma~\ref{#1}}

\newcommand\appref[1]{Appendix~\ref{#1}}

\begin{document}
\pagestyle{empty}
\setcounter{page}{0}
\begin{flushright}
\sf solv-int/9807008 \\
    UTMS 98-28 \\
    PDMI 15/98
\end{flushright}
\vskip1cm
\begin{center}
\Large\bf
Separation of Variables \\
in the Elliptic Gaudin Model
\end{center}
\vskip1cm
\begin{center}
Evgueni K. Sklyanin\\
St.\ Petersburg Branch \\ of the Steklov Mathematical Institute,\\
Fontanka 27, St.\ Petersburg 191011, Russia,\\
{\tt sklyanin@pdmi.ras.ru}
\vskip5mm
Takashi Takebe\\
Department of Mathematical Sciences,\\
the University of Tokyo,\\
Komaba 3-8-1, Meguro-ku,\\
Tokyo, 153-8914, Japan.\\
{\tt takebe@ms.u-tokyo.ac.jp}
\vskip1cm
Running title: {\bf Elliptic Gaudin Model}
\end{center}

\vskip1.5cm
{\bf Abstract.}
For the elliptic Gaudin model (a degenerate case of XYZ integrable spin chain)
a separation of variables is constructed in the classical case. The
corresponding separated coordinates are obtained as the poles of a
suitably normalized Baker-Akhiezer function. The classical results
are generalized to the quantum case where the kernel of separating integral
operator is constructed. The simplest one-degree-of-freedom case is studied
in detail.

\newpage
\section{Introduction}
\pagestyle{plain}

The quantum elliptic (or XYZ) Gaudin model was introduced in
\cite{gaudin:76}, see
also \cite{gaudin:83}, as a limiting case of the integrable XYZ spin
chain \cite{baxter:book}. The commuting Hamiltonians $H_n$ of the model are
expressed as quadratic combinations of $sl_2$ spin operators.
Determining the spectrum of $H_n$ turned out to be a difficult problem
like the original XYZ spin chain. Let us list the known facts related to
this problem.

\begin{itemize}
\item
A solution by means of Algebraic Bethe Ansatz
has been obtained only recently \cite{skl-tak:96}.
See also \cite{b-ls-p:98}.

\item
As shown in \cite{skl:87}, in the SU(2)-invariant, or XXX, or rational,
case the spectrum
and the eigenfunctions of the model can be found via an alternative method,
Separation of Variables, see also the survey \cite{skl:95}.

\item
In \cite{fre:95}  the separation of variables in the rational Gaudin model
\cite{skl:87} was interpreted as a geometric Langlands correspondence.

\item
In \cite{e-f-r:98} a separation of variables was constructed for
the elliptic Gaudin-Calogero model which is closely related to the
XYZ Gaudin model, though the separation of variables for the former one
is much simpler.

\item
The results of \cite{fre:95} and \cite{e-f-r:98} are based on the
interpretation of the corresponding Gaudin models
as conformal field theoretical models (Wess-Zumino-Witten models).
The corresponding
interpretation of the XYZ Gaudin model was obtained in
\cite{kur-tak:97}, but the conformal field theoretical model
corresponding to the XYZ Gaudin model turned out to be so complicated that
writing
down the geometric Langlands correspondence for this system, following
\cite{fre:95}, is not easy.
\end{itemize}

The main task of the present paper is to present a construction of
separated variables for the XYZ Gaudin model both in the classical and
quantum cases.
The paper is organized as follows. After giving a detailed description of the
XYZ Gaudin model in \secref{definitions}, we proceed, in
\secref{classical-sov}, with the 
classical case and, following the general philosophy of \cite{skl:95},
construct the separated coordinates as the poles of an appropriately
normalized Baker-Akhiezer function. The corresponding eigenvalues of the Lax
matrix are then shown to provide the canonically conjugated momenta.
The whole construction is a simplified version of the one used in
\cite{skl:86}.

The quantum case is considered in \secref{quantum}.
The separating classical canonical
transformation is replaced by an integral operator $\sov$. We write down a
system of differential equations for the kernel of $\sov$ and show
that it is integrable. The resulting integral operator $\sov$
intertwines the original and
the separated variables and provides, respectively, a Radon-Penrose
transformation of the corresponding D-modules. The quantization constructed
is a formal one, since we do not study the transformations of the functional
spaces of quantum states, leaving it for a further study. A detailed study
of the spectral problem is given in the simplest case only: $N=1$
(\secref{N=1}). 
We show that the corresponding separated equation is none other than
the (generalized) Lam\'{e} equation.
Two appendices contain, respectively, a list of properties of elliptic
functions, and the formulas describing a realization of finite-dimensional
representations of $sl_2(\Complex)$ on the elliptic curve which are used
throughout the paper.

One of the authors (E.S.) is grateful to the Department of Applied
Mathematics, University of Leeds, UK, where the most part of the paper was
written, for hospitality and acknowledges the support of EPSRC.  The other
one (T.T.) is grateful to the Department of Mathematics, University of
California at Berkeley, USA, where some part of the work was done, for
hospitality. He also acknowledges the support of Postdoctoral Fellowship
for Research abroad of JSPS. Special thanks are to Benjamin Enriquez and
Vladimir Rubtsov who gave us their article \cite{e-f-r:98} before
publication and explained details.

\section{Description of the model}
\label{definitions}
\setcounter{equation}{0}
Let us recall the definition of the XYZ Gaudin model,
following \cite{skl-tak:96}. The elementary Lax operator $L(u)$ of the model
depending on a complex parameter $u$ (spectral parameter)
is given by
\begin{equation}
    {L} (u) =
    \frac12 \sum_{a = 1}^3 w_a(u) S^a \tensor \sigma^a
    =
    \pmatrix{
    \calA(u) & \calB(u) \cr
    \calC(u) & -\calA(u)
    }.
\label{def:L}
\end{equation}
Here $\sigma^a$ are the Pauli matrices,
\begin{equation}
    w_1(u)
    = \frac{\theta'_{11}}{\theta_{10}}
      \frac{\theta_{10}(u)}{\theta_{11}(u)},
    \quad
    w_2(u)
    = \frac{\theta'_{11}}{\theta_{00}}
      \frac{\theta_{00}(u)}{\theta_{11}(u)},
    \quad
    w_3(u)
    = \frac{\theta'_{11}}{\theta_{01}}
      \frac{\theta_{01}(u)}{\theta_{11}(u)},
\label{def:w-a}
\end{equation}
where
$\theta_{\a\b}(u) = \theta_{\a\b}(u;\tau)$, $\theta_{\a\b}=\theta_{\a\b}(0)$,
$\theta'_{11}= d/du (\theta_{11}(u))|_{u=0}$, (see \appref{notations})
and $S^a$ are generators of the Lie algebra $sl_2(\Complex)$:
$$ 
    [S^a, S^b] = i S^c.
$$ 
Hereafter $(a,b,c)$ denotes a cyclic permutation of $(1,2,3)$.  Note that
$\calA$, $\calB$, $\calC$ are holomorphic except at
$u \in \Integer+\tau\Integer$, where these operators have poles of first
order.

Introducing the notation $\one L:=L\otimes\id_2$ and $\two L:=\id_2\otimes L$,
where $\id_2$ is the unit operator in $\Complex^2$,
one can establish the commutation relation
\begin{equation}
    [\one{L}(u), \two{L}(v)]
    =
    [r(u-v), \one{L}(u) + \two{L}(v)],
\label{comm-rel:L}
\end{equation}
where $r(u)$ is a classical $r$ matrix defined by
\begin{equation}
    r(u) = -\frac{1}{2} \sum_{a=1}^3 w_a(u) \s^a\tensor \s^a.
\label{def:r}
\end{equation}
The $r$ matrix behaves as $-\frac{1}{u}(\calP - \frac{1}{2})+O(u^{-3})$
when $u \to 0$. Here $\calP$ is the permutation operator:
$\calP (x\tensor y) = y \tensor x$.
Explicitly, in the natural basis in $\Complex^2\otimes\Complex^2$,
\begin{equation}
    r(u) = \pmatrix{
                    a(u) & 0    & 0    & d(u) \cr
                    0    & b(u) & c(u) & 0 \cr
                    0    & c(u) & b(u) & 0 \cr
                    d(u) & 0    & 0    & a(u)
            },
\label{r-matrix}
\end{equation}
where
$$ 
    a(u) = -b(u) = - \frac{w_3(u)}{2}, \quad
    c(u) = -\frac{w_1(u) + w_2(u)}{2}, \quad
    d(u) = -\frac{w_1(u) - w_2(u)}{2}.
$$ 

Since $w_a(u)$ are quasiperiodic in $u$ because of
(\ref{period:theta}):
\begin{equation}
\begin{array}{rcrcr}
    w_1(u) &=& w_1(u+1) &=& -w_1(u+\tau), \\
    w_2(u) &=&-w_2(u+1) &=& -w_2(u+\tau), \\
    w_3(u) &=&-w_3(u+1) &=&  w_3(u+\tau),
\end{array}
\end{equation}
the ${L}$ operator (\ref{def:L}) has the following quasiperiodicity:
\begin{equation}
    {L}(u + 1   ) = \sigma^1 {L}(u) \sigma^1, \qquad
    {L}(u + \tau) = \sigma^3 {L}(u) \sigma^3.
\label{L:period}
\end{equation}

Let $\ell_n$ ($n=1,\ldots, N$) be half integers.  The total Hilbert space
of the model is $V = \bigotimes_{n=1}^N V_n$, where
$V_n \simeq V^{(\ell_n)}$ and $V^{(\ell)}$ is a spin $\ell$ representation
space of $sl_2$:
\begin{equation}
    \rho^{\ell}: sl_2(\Complex) \to \End_{\Complex}(V^{(\ell)}),
\qquad  V^{(\ell)} \simeq \Complex^{2\ell+1}.
\label{spin-l-rep}
\end{equation}
The generating function of the integrals of motion is
\begin{equation}
    \hat\tau(u) = \frac12 \trace {T}^2(u),
\label{def:tau}
\end{equation}
where the matrix ${T}(u)$ is constructed as the sum of elementary Lax
operators (\ref{def:L})
\begin{equation}
    {T}(u) = \sum_{n=1}^N {L}_n (u - z_n)
             = \pmatrix{
                A(u) & B(u) \cr
                C(u) & -A(u)
               }.
\label{def:T}
\end{equation}
Here $z_n$ are mutually distinct complex parameters,
\begin{eqnarray}
    {L}_n(u)&=&
    \frac12
    \sum_{a=1}^3 w_a(u) S^a_n\tensor \s^a \\
    S^a_n &=& \id_{V_1}\tensor \dots \tensor \id_{V_{n-1}}
                              \tensor \rho^{\ell_n}(S^a) \tensor
                             \id_{V_{n+1}}\tensor \dots \tensor \id_{V_N}.
\end{eqnarray}
By virtue of the commutation relations (\ref{comm-rel:L}) the
operator ${T}(u)$ satisfies the same commutation relations
\begin{equation}
    [\one{T}(u), \two{T}(v)]
    =
    [r(u-v), \one{T}(u) + \two{T}(v)],
\label{comm-rel:T}
\end{equation}
which implies the commutativity of $\hat\tau(u)$:
\begin{equation}
    [\hat\tau(u), \hat\tau(v)]=0,
\label{comm-rel:tau}
\end{equation}

Operator $\hat\tau(u)$ is explicitly written down as follows:
\begin{equation}
    \hat\tau(u)
    = \sum_{n=1}^N \wp_{11}(u-z_n) \ell_n(\ell_n+1)
      + \sum_{n=1}^N H_n \zeta_{11}(u-z_n) + H_0.
\label{tau}
\end{equation}
Here $\wp_{11}$, $\zeta_{11}$ are normalized Weierstra{\ss} functions
defined by (\ref{def:zeta-p-11}) and
\begin{eqnarray}
    H_n &=& \frac12 \sum_{m \neq n} \sum_{a=1}^3 w_a(z_n - z_m) S_n^a S_m^a,
\nonumber\\
    H_0 &=& \sum_{n,m=1}^N \sum_{a=1}^3 Z_a(z_n - z_m) S_n^a S_m^a
\label{def:H-n}
\end{eqnarray}
are integrals of motion, where
\begin{equation}
    Z_1(t) = \frac{\theta'_{11}}{4\theta_{10}}
             \frac{\theta'_{10}(t)}{\theta_{11}(t)}, \qquad
    Z_2(t) = \frac{\theta'_{11}}{4\theta_{00}}
             \frac{\theta'_{00}(t)}{\theta_{11}(t)}, \qquad
    Z_3(t) = \frac{\theta'_{11}}{4\theta_{01}}
             \frac{\theta'_{01}(t)}{\theta_{11}(t)}.
\label{def:Z-a}
\end{equation}

Note that the integrals of motion $H_n$ ($n=0, \dots, N$) appear as
coefficients of the elliptic Knizhnik-Zamolodchikov equations in
\cite{kur-tak:97}. Our expression (\ref{def:H-n}) for $H_0$ differs from
that given in \cite{skl-tak:96}
because of different normalization of the $\wp$ and $\zeta$
functions.

The classical Gaudin model is obtained if we replace all the commutators
with the Poisson brackets, e.g.
\begin{equation}
    \{ \one{T}(u), \two{T}(v) \}
    =
    [r(u-v), \one{T}(u) + \two{T}(v)],
\label{poisson:T}
\end{equation}
instead of (\ref{comm-rel:T}). The spin variables $S^a$
satisfy, respectively, the Poisson commutation relations
$[S^a, S^b] = i S^c$ and are subject to the constraint
$\sum_{a=1}^3 (S^a)^2=\ell^2$.

\section{Classical separation of variables}
\label{classical-sov}
\setcounter{equation}{0}

According to the recipe in \cite{skl:95}, the separated coordinates $x_n$
should be constructed as the poles of a suitably normalized
Baker-Akhiezer function (eigenvector of Lax matrix $T(u)$). The corresponding
canonically conjugated variables should appear then as the corresponding
eigenvalues of $T(x_n)$.
Instead of choosing a normalization, we shall rather speak of a choice of a
gauge transformation $M$ of $T(u)$. The separated coordinates $x_n$ will
be obtained then as the zeros
of the off-diagonal element $\tilde B(u)$ of the twisted matrix
$\tilde T=M^{-1}TM$.

The classical XYZ Gaudin model is a degenerate case of
the classical lattice Landau-Lifshits
equation for which a separation of variables has been constructed in
\cite{skl:86}, see also a discussion in \cite{skl:95}. Here we use
essentially the same gauge transformation $M(u)$
as in \cite{skl:86}, and our calculations represent a revised and simplified
version of those in \cite{skl:86}.

\subsection{Gauge transformation}
\label{gauge-transformation}

Let $M(u;\utilde)$ be the following $2 \times 2$ matrix
\begin{equation}
    M(u;\utilde) :=
      \pmatrix{
      - \theta_{01}\left(\frac{u-\utilde}{2};\frac{\tau}{2} \right) &
      - \theta_{01}\left(\frac{u+\utilde}{2};\frac{\tau}{2} \right)\cr
        \theta_{00}\left(\frac{u-\utilde}{2};\frac{\tau}{2} \right) &
        \theta_{00}\left(\frac{u+\utilde}{2};\frac{\tau}{2} \right)
      },
\label{def:M}
\end{equation}
where $u$ and $\utilde$ are (possibly dynamical) parameters.
(This matrix appears also in the context of the algebraic Bethe Ansatz. See
\cite{takh-fad:79}, \cite{skl-tak:96}.)
A twisted $L$-operator $\tilde L(u,v;\tilde u)$ depending
on a parameter $\utilde$ is defined by
\begin{equation}
    \tilde {L}(u,v;\utilde)
    = \pmatrix{
      \tilde\calA(u,v;\utilde) &   \tilde\calB(u,v;\utilde) \cr
      \tilde\calC(u,v;\utilde) & - \tilde\calA(u,v;\utilde)
      }
    := M^{-1}(u;\utilde) {L}(u-v) M(u;\utilde).
\label{def:tilde-L}
\end{equation}
Likewise we define the twisted Lax matrix by
\begin{equation}
    \tilde {T}(u;\utilde)
    = \pmatrix{
      \tilde A(u;\utilde) &  \tilde B(u;\utilde) \cr
      \tilde C(u;\utilde) & -\tilde A(u;\utilde)
      }
    := M^{-1}(u;\utilde) {T}(u) M(u;\utilde).
\label{def:tilde-T}
\end{equation}

Note that $M(u;\utilde)$ has the quasiperiodicity because of
(\ref{period:theta}):
\begin{eqnarray}
    M(u+1   ;\utilde) &=& - \sigma_1 M(u;\utilde), \nonumber \\
    M(u+\tau;\utilde) &=&  e^{- \pi i (u + \tau/2)}
              \sigma_3 M(u;\utilde) \exp({\pi i \utilde\sigma_3}).
\label{M:period}
\end{eqnarray}

These formulae together with (\ref{L:period}) imply that the function
$\tilde B(u,v;\utilde)$ has the following quasiperiodicity properties:
\begin{equation}
    \tilde B(u+1   ;\utilde) = \tilde B(u;\utilde), \qquad
    \tilde B(u+\tau;\utilde) = e^{-2\pi i \utilde} \tilde B(u;\utilde).
\label{B-period}
\end{equation}
Hence by a standard argument in the theory of elliptic functions
(see \cite{whit-wat}), we have
\begin{equation}
    \deg(\div(\tilde B(u))) = 0, \qquad
    - \utilde + \sum (\mult_y \div(\tilde B(u)))\, y
    \in \Integer+\tau\Integer,
\label{tildeB-div}
\end{equation}
where $\mult_y \div(\tilde B(u))$ is the multiplicity of a divisor
$[y]$ in the divisor $\div(\tilde B(u))$.  By the definition
(\ref{def:tilde-T}), operator $\tilde B(u;\utilde)$ is holomorphic
except at poles of $A(u)$, $B(u)$, $C(u)$, i.e., $u=z_n$ ($n=1, \dots,
N$), and at zeros of $\det M(u;\utilde)$, i.e., $u=0$ modulo
$\Integer+\tau\Integer$:
\begin{equation}
    \div(\tilde B(u)) \geq - \left(\sum_{n=1}^N [z_n] + [0] \right)
    \pmod{\Integer+\tau\Integer}.
\label{tildeB-pole}
\end{equation}
Thus (\ref{tildeB-pole}) and (\ref{tildeB-div}) imply that there are
($N+1$) points $x_0, \dots, x_N$ such that
\begin{equation}
    \div(\tilde B(u)) \equiv \sum_{j=0}^N [x_j]
                      - \left(\sum_{n=1}^N [z_n] + [0]\right)
                      \pmod{\Integer+\tau\Integer},
\label{def:x-j}
\end{equation}
and
\begin{equation}
    \sum_{j=0}^N x_j \equiv \sum_{n=1}^N z_n - \utilde
    \pmod{\Integer+\tau\Integer}.
\label{sum-x-j}
\end{equation}

Let us fix the parameter $\utilde$ by the condition that one of $x_j$,
for example $x_0$, is a constant $\xi$. Note that $\utilde$ becomes then
a dynamical variable. Thus we have
\begin{equation}
    \tilde B(x_j; \utilde) =
    \tilde B(u = \xi; \utilde) = 0.
\label{zero-of-tilde-B}
\end{equation}

Dynamical variables $x_1, \ldots, x_N$ are (classically) separated
coordinates of the system as we will see below.

\subsection{Poisson commutation relations and classical separation of
variables}
\label{poisson-rel}

The main purpose of this subsection is to prove the following commutation
relations.

\begin{thm}
\label{canonical-poisson-relation}
Generically the dynamical variables $x_j$ and $-\tilde A(x_j)$
have the canonical Poisson brackets:
\begin{enumerate}
\renewcommand{\labelenumi}{{\rm (\roman{enumi})}}
\item
$\{x_i, x_j\} = 0$ for all $i, j = 1, \ldots, N$.

\item
$\{-\tilde A(x_i), -\tilde A(x_j)\} = 0$ for all $i, j = 1, \ldots, N$.

\item
$\{-\tilde A(x_i), x_j\} = \delta_{i,j}$ for all $i, j = 1, \ldots, N$.

\end{enumerate}
\end{thm}

To prove the theorem, we follow the argument of \cite{skl:86}.
First let us introduce several
notations. Define the matrices $\hat A$, $\hat B$, $\hat C$, $\hat D$ as
\begin{equation}
    \hat A:= \pmatrix{ 1 & 0 \cr 0 & 0 },\quad
    \hat B:= \pmatrix{ 0 & 1 \cr 0 & 0 },\quad
    \hat C:= \pmatrix{ 0 & 0 \cr 1 & 0 },\quad
    \hat D:= \pmatrix{ 0 & 0 \cr 0 & 1 }.
\label{def:hat-A,B,C,D}
\end{equation}
Gauge transformation of them are defined as follows:
\begin{equation}
\begin{array}{rlrl}
    \hat A(u;\utilde) &:= M(u;\utilde)\hat A M(u;\utilde)^{-1}, &\quad
    \hat B(u;\utilde) &:= M(u;\utilde)\hat B M(u;\utilde)^{-1}, \\
    \hat C(u;\utilde) &:= M(u;\utilde)\hat C M(u;\utilde)^{-1}, &\quad
    \hat D(u;\utilde) &:= M(u;\utilde)\hat D M(u;\utilde)^{-1}.
\end{array}
\label{def:hat-A,B,C,D(u)}
\end{equation}
Bracket $\langle , \rangle$ is the standard inner product of the
$2\times2$ matrices:
\begin{equation}
    \langle X, Y \rangle = \tr XY.
\label{def:inner-prod}
\end{equation}
When $X(u)$ is a variable depending on the spectral parameter $u$, we will
denote $X(x_i)$ by $X_i$ for brevity. For example,
$$
    (\partial_u \langle \hat C T \rangle)_i
    = \left. \frac{\partial}{\partial u}\right|_{u=x_i}
     \tr(\hat C(u;\utilde) T(u)).
$$

The following statement is proved by the same argument as
in the proof of Theorem in \S2 of \cite{skl:86}.

\begin{lem}
\label{poisson-auxiliary}
For any dynamical variable $X$,
\begin{equation}
    \{X, \tilde u\} =
    - \frac{\langle \hat C_0 \{X, T\}_0 \rangle}
           {\langle \partial_{\utilde}\hat C_0 T_0 \rangle},
\label{poisson-com-with-utilde}
\end{equation}
and
\begin{equation}
    \{X, x_j\} =
    \frac
    {\langle \hat C_0 \{X, T\}_0 \rangle
     \langle \partial_{\utilde}\hat C_j T_j \rangle
     -
     \langle \hat C_j \{X, T\}_j \rangle
     \langle \partial_{\utilde}\hat C_0 T_0 \rangle}
    {(\partial_u \langle \hat C,T \rangle)_j
     \langle \partial_{\utilde}\hat C_0 T_0 \rangle}.
\label{poisson-com-with-x-j}
\end{equation}
\end{lem}

We also need the formula for the twisted $r$ matrix.
\begin{lem}
\label{twisted-r}
Define
\begin{equation}
    \tilde r(u,v;\utilde) :=
    \one M(u;\utilde)^{-1} \two M(v;\utilde)^{-1} r(u-v)
    \one M(u;\utilde)      \two M(v;\utilde),
\label{def:twisted-r}
\end{equation}
which we call the {\em twisted $r$ matrix} and let
$\tilde r_{ij}(u,v;\utilde)$ be its $(i,j)$ element. Then it has the
following form.
\begin{eqnarray}
\lefteqn{\tilde r_{11}(u,v;\utilde) = - \tilde r_{22}(u,v;\utilde) =
    - \tilde r_{33}(u,v;\utilde) =   \tilde r_{44}(u,v;\utilde) =  }
    \nonumber\\
    &\hspace{3cm}=& -\frac12 \left(
      \frac{\theta_{11}'(u-v)}{\theta_{11}(u-v)}
    - \frac{\theta_{11}'(u  )}{\theta_{11}(u  )}
    + \frac{\theta_{11}'(v  )}{\theta_{11}(v  )}
    \right),
\label{twisted-r-diagonal}
\\
\lefteqn{\tilde r_{12}(u,v;\utilde) = - \tilde r_{13}(v,u;\utilde) =
     - \tilde r_{21}(u,v;-\utilde)=   \tilde r_{31}(v,u;-\utilde) = }
    \nonumber\\
\lefteqn{\hspace{1cm}
  =\tilde r_{24}(v,u;\utilde) = - \tilde r_{34}(u,v;\utilde) =
     - \tilde r_{42}(v,u;-\utilde)=   \tilde r_{43}(u,v;-\utilde) = }
    \nonumber\\
    &\hspace{3cm}=&\frac{-\theta'_{11}\theta_{11}(v+\utilde)}
           {2\theta_{11}(\utilde)\theta_{11}(v)},
\label{twisted-r-12}
\\
\lefteqn{
    \tilde r_{14}(u,v:\utilde) = \tilde r_{41}(u,v;\utilde) = 0,
    }
\label{twisted-r-14}
\\
\lefteqn{
    \tilde r_{23}(u,v;\utilde) = \tilde r_{32}(u,v;-\utilde) =
    \frac{-\theta'_{11} \theta_{11}(u-v+\utilde)}
         { \theta_{11}(u-v) \theta_{11}(\utilde)}.
    }
\label{twisted-r-23}
\end{eqnarray}

\end{lem}

{\it Proof\/} is given by a direct computation. For example, we have
formulae like
\begin{eqnarray*}
    M(u,\utilde)^{-1} \sigma_1 M(u,\utilde)
    &=&
    \frac1{\theta_{11}(u) \theta_{11}(\utilde)}
    \pmatrix{
      \theta_{10}(u)\theta_{10}(\utilde) & \theta_{10}\theta_{10}(u+\utilde)
     \cr
     -\theta_{10}\theta_{10}(u-\utilde) &-\theta_{10}(u)\theta_{10}(\utilde)
    },
\\
    M(u,\utilde)^{-1} (i\sigma_2) M(u,\utilde)
    &=&
    \frac1{\theta_{11}(u) \theta_{11}(\utilde)}
    \pmatrix{
      \theta_{00}(u)\theta_{00}(\utilde) & \theta_{00}\theta_{00}(u+\utilde)
     \cr
     -\theta_{00}\theta_{00}(u-\utilde) &-\theta_{00}(u)\theta_{00}(\utilde)
    },
\\
    M(u,\utilde)^{-1} \sigma_3 M(u,\utilde)
    &=&
    \frac1{\theta_{11}(u) \theta_{11}(\utilde)}
    \pmatrix{
     -\theta_{01}(u)\theta_{01}(\utilde) &-\theta_{01}\theta_{01}(u+\utilde)
     \cr
      \theta_{01}\theta_{01}(u-\utilde) & \theta_{01}(u)\theta_{01}(\utilde)
    },
\end{eqnarray*}
which follow from the addition theorems (cf.~\cite{mum}, p.~20, p.~22) and
the Landen transformation (cf.\ \cite{whit-wat} \S21.52) of theta
functions. Substituting them in the definition of $\tilde r$
(\ref{def:twisted-r}) and using the addition theorems again, we can prove
the lemma. Q.E.D.

\medskip
{\it Proof\/} of \thmref{canonical-poisson-relation}.

Using the formulae (\ref{poisson-com-with-utilde}) and
(\ref{poisson-com-with-x-j}), we have
\begin{eqnarray}
\lefteqn{    \{x_j, x_k\} =
    \frac{1}{(\partial_u\langle\hat C T\rangle)_j
             (\partial_u\langle\hat C T\rangle)_k}\times  }
    \nonumber \\
&&\times
    \Bigl[
    \frac{\langle\partial_\utilde \hat C_j T_j\rangle
          \langle\partial_\utilde \hat C_k T_k\rangle}
         {\langle\partial_\utilde \hat C_0 T_0\rangle^2}
    \langle
     \one{\hat C}_0 \two{\hat C}_0 \{\one T, \two T\}_{00}
    \rangle
    - \frac{\langle\partial_\utilde \hat C_k T_k\rangle}
         {\langle\partial_\utilde \hat C_0 T_0\rangle}
    \langle
     \one{\hat C}_j \two{\hat C}_0 \{\one T, \two T\}_{j0}
    \rangle \nonumber \\
&& - \frac{\langle\partial_\utilde \hat C_j T_j\rangle}
         {\langle\partial_\utilde \hat C_0 T_0\rangle}
    \langle
     \one{\hat C}_0 \two{\hat C}_k \{\one T, \two T\}_{0k}
    \rangle
    + \langle\one{\hat C}_j \two{\hat C}_k
         \{\one T, \two T\}_{jk}\rangle
    \Bigr].
\label{x-j-x-k:tmp}
\end{eqnarray}
Therefore computation of $\{x_j,x_k\}$ reduces to computation of
$\langle\one{\hat C}_j \two{\hat C}_k \{\one T, \two T\}_{jk}\rangle$.

As in Appendix B of \cite{skl:86}, we have
\begin{equation}
    \langle\one{\hat \Phi}_j \two{\hat \Psi}_k
           \{\one T, \two T\}_{jk}\rangle
    =
    \tr_1 \tr_2 ([\one{\hat \Phi} \two{\hat \Psi}, \tilde r(x_j, x_k)]
    (\one{\tilde T}(x_j;\utilde) + \two{\tilde T}(x_k;\utilde) ) ),
\label{Phi-Psi-TT}
\end{equation}
for any $\Phi, \Psi = A, B, C, D$. Substituting $\Phi = \Psi = C$ and
using (\ref{twisted-r-14}), we have
$\langle\one{\hat C}_j \two{\hat C}_k \{\one T, \two T\}_{jk} \rangle
= 0$. Thus (\ref{x-j-x-k:tmp}) implies that $\{x_j, x_k\}=0$.

A direct consequence of this is $\{x_j, \utilde\}=0$, which follows
from (\ref{sum-x-j}).  Using these results and
\lemref{poisson-auxiliary}, we have for $j\neq k$
\begin{equation}
    \{\tilde A(x_j), x_k\}
    =
    \frac{
    \langle \partial_{\utilde}\hat C_k T_k \rangle
    \langle
     \one{\hat A}_j \two{\hat C}_0 \{\one T, \two T\}_{j0}
    \rangle
    -
    \langle \partial_{\utilde}\hat C_0 T_0 \rangle
    \langle
     \one{\hat A}_j \two{\hat C}_k \{\one T, \two T\}_{jk}
    \rangle}
    {(\partial_u \langle \hat C,T \rangle)_k
      \langle \partial_{\utilde}\hat C_0 T_0 \rangle}.
\label{A(x-j)-x-k:tmp}
\end{equation}
Hence we need to know
$\langle\one{\hat A}_j \two{\hat C}_k \{\one T, \two T\}_{jk} \rangle$
and $\langle \partial_{\utilde}\hat C_k T_k \rangle$. The former can
be computed by (\ref{Phi-Psi-TT}) and (\ref{twisted-r-14}) and we have
\begin{equation}
    \langle
     \one{\hat A}_j \two{\hat C}_k \{\one T, \two T\}_{jk}
    \rangle
    = -2 \tilde r_{12}(x_j,x_k;\utilde) \tilde A_k.
\label{A-C-TT}
\end{equation}
The factor $\langle \partial_{\utilde}\hat C_k T_k \rangle$ is
computed as follows:
\begin{eqnarray}
    \lefteqn{
    \langle \partial_{\utilde}\hat C_k T_k \rangle
    =
    \langle
     [M(x_k;\utilde)^{-1} \partial_{\utilde}M(x_k;\utilde), \hat C]
     \tilde T(x_k;\utilde)
    \rangle =
    }
    \nonumber\\
&\hspace{5cm}=&
    - \frac{\theta'_{11} \theta_{11}(x_k+\utilde)}
           {\theta_{11}(x_k) \theta_{11}(\utilde)} \tilde A_k.
\label{der-utilde-CT}
\end{eqnarray}
Substituting (\ref{A-C-TT}), (\ref{der-utilde-CT}) and
(\ref{twisted-r-12}) into (\ref{A(x-j)-x-k:tmp}), we have
$\{\tilde A(x_j), x_k\} = 0$ for $j\neq k$.

The proof of $\{\tilde A(x_j), \tilde A(x_k)\} = 0$ is done in a
similar way. In addition to the formulae we have shown above, we need 
\begin{eqnarray}
    \langle
     \one{\hat A}_j \two{\hat A}_k \{\one T, \two T\}_{jk}
    \rangle
    &=& \tilde r_{13}(x_j, x_k;\utilde) \tilde C_j
      + \tilde r_{12}(x_j, x_k;\utilde) \tilde C_k,
\label{A-A-TT}
\\
    \langle \partial_{\utilde}\hat A_k T_k \rangle
    &=&
    \frac{-\theta'_{11} \theta_{11}(x_k+\utilde)}
         {2\theta_{11}(x_k) \theta_{11}(\utilde)} \tilde C_k.
\label{der-utilde-AT}
\end{eqnarray}

Proof of the remaining equation $\{\tilde A(x_j), x_j\} = -1$ requires
a special care, since the $r$ matrix $r(u)$ diverges at $u=0$. Instead
of (\ref{A-C-TT}), we use 
\begin{eqnarray}
    \langle
     \one{\hat A}_j \two{\hat C}_j \{\one T, \two T\}_{jj}
    \rangle
    &=& -2 \tilde r_{12}(x_j, x_j;\utilde) \tilde A_j
    - \lim_{u\to x_j} 
      \tilde r_{32}(u,x_j;\utilde) \tilde B(u;\utilde)
    \nonumber\\
    &=& -2 \tilde r_{12}(x_j,x_j;\utilde) \tilde A_j
       + (\partial_u \tilde B)_j.
\label{(A-C-TT)jj}
\end{eqnarray}
Noting 
$(\partial_u \langle \hat C,T \rangle)_j = (\partial_u \tilde B)_j$ and
substituting (\ref{(A-C-TT)jj}) and (\ref{der-utilde-CT}) into
(\ref{A(x-j)-x-k:tmp}), we have $\{\tilde A(x_j), x_j\} = -1$.
Q.E.D.

\medskip
Since $\tilde B$ is zero at $u=x_j$, dynamical variable
$X_j:=- \tilde A(x_j)$ is an eigenvalue of ${T}(x_j)$:
\begin{eqnarray}
    \tilde {T}(x_j)
    \left( \begin{array}{c} 0 \\ 1 \end{array} \right)
    &=&
    X_j
    \left( \begin{array}{c} 0 \\ 1 \end{array} \right),
\\
    {T}(x_j)
    \left(
    \begin{array}{c} 
      - \theta_{01}\left(\frac{x_j+\utilde}{2};\frac{\tau}{2} \right) \cr
        \theta_{00}\left(\frac{x_j+\utilde}{2};\frac{\tau}{2} \right)
    \end{array}
    \right)
    &=&
    X_j
    \left(
    \begin{array}{c} 
      - \theta_{01}\left(\frac{x_j+\utilde}{2};\frac{\tau}{2} \right) \cr
        \theta_{00}\left(\frac{x_j+\utilde}{2};\frac{\tau}{2} \right)
    \end{array}
    \right)
\end{eqnarray}
Thus if we define the {\em characteristic polynomial\/} by
\begin{equation}
    W(z,u) := \det (z - {T}(u)),
\label{def:char-poly}
\end{equation}
each pair of dynamical variables $(x_j, X_j)$ satisfies an
equation
\begin{equation}
    W(X_j, x_j) = 0,
\label{cl-separated-eq}
\end{equation}
for $j=1, \dots, N$. Therefore, following the definition in \cite{skl:95},
canonical variables $(x_1, \dots, x_N; X_1, \dots, X_N)$ are {\em
separated variables} of the classical ellipitic Gaudin model.

\section{Quantum System: general case}
\label{quantum}
\setcounter{equation}{0}

We return now to the quantum elliptic Gaudin model and construct the quantum
separation of variables.
The special case $N=1$ is considered in the next section, \secref{N=1}.

\subsection{Kernel function}

Suppose that the representation space $V_n = V^{\ell_n}$ (\ref{spin-l-rep})
is realized as a space of functions on a certain space with coordinate
$y_n$ and that the operators $S^a$ are differential operators on, e.g.,
polynomials or elliptic functions. The separating operator $\sov$ is
expressed as an integral operator
\begin{equation}
   \sov f(x_1,\ldots,x_N) =
   \int dy_1 \cdots dy_N\, \Phi(x_1,\ldots,x_N|y_1,\ldots, y_N)
      f(y_1,\ldots,y_N),
\label{def:sov}
\end{equation}
which maps a function of $(y_1,\ldots,y_N)$ in $V_1\tensor \cdots \tensor
V_N$ to a function of $N$-variables $x_i$ on the elliptic curve
$\Complex/\Integer+\tau\Integer$.

Let us define the operator $X_i$ as follows:
\begin{equation}
    X_i := \frac{\partial}{\partial x_i} - \Lambda(x_i),
    \qquad
    \Lambda(x) = 
    \sum_{n=1}^N \ell_n \frac{\theta'_{11}(x-z_n)}{\theta_{11}(x-z_n)}.
\label{def:Xi}
\end{equation}

\begin{lem}
\label{consistency-kernel}
The following system of partial differential equations satisfies the
Fro\-be\-nius integrability condition.
\begin{eqnarray}
    \tilde B^\ast(x_i;\utilde) \Phi &=& 0, 
    \qquad i = 1,\ldots, N,
\label{B-Phi}
\\
    (X_i + {\tilde A}^\ast(x_i;\utilde)) \Phi &=& 0,
    \qquad i = 1,\ldots, N,
\label{A-Phi}
\end{eqnarray}
where $P^\ast$ is the (formal) adjoint of a differential operator $P$
with respect to $(y_1,\ldots,\allowbreak y_N)$ and we set
\begin{equation}
    \utilde = \sum_{n=1}^N z_n - \sum_{j=0}^N x_j
\label{fix:utilde}
\end{equation}
for a certain constant $x_0=\xi$.
\end{lem}

{\it Proof\/}.
This is a consequence of the commutation relation (\ref{comm-rel:T}). By
multiplying $\one M(u;\utilde) \two M(v;\utilde)$ from the right and its
inverse from the left, we have
\begin{equation}
    [\one{\tilde{T}}(u;\utilde), \two{\tilde{T}}(v;\utilde)]
    =
    [\tilde r(u,v;\utilde), 
    \one{\tilde{T}}(u;\utilde) + \two{\tilde{T}}(v;\utilde)].
\label{comm-rel:twisted-T}
\end{equation}
Note that $\utilde$ is {\em not} a dynamical variable in contrast to that
in \secref{classical-sov}.

In order to show the consistency of the equations (\ref{B-Phi}) for
$i$ and for $j$, we prove that
$[\tilde B^\ast(x_i;\utilde), \tilde B^\ast(x_j;\utilde)]$ is expressed as
a linear combination of $\tilde B^\ast(x_i;\utilde)$ and 
$\tilde B^\ast(x_j;\utilde)$.

Since the formal adjoint is an algebra anti-isomorphism,
$(PQ)^\ast = Q^\ast P^\ast$, we have
\begin{equation}
    [\tilde B^\ast(x_i;\utilde), \tilde B^\ast(x_j;\utilde)]
    =
    [\tilde B(x_j;\utilde), \tilde B(x_i;\utilde)]^\ast.
\label{B-B:tmp}
\end{equation}
The (1,4)-element of (\ref{comm-rel:twisted-T}) gives
\begin{equation}
    [\tilde B(u;\utilde), \tilde B(v;\utilde)]
    =
    2(\tilde r_{12}(u,v;\utilde) \tilde B(u) - 
      \tilde r_{12}(v,u;\utilde) \tilde B(v))
\label{B-B:tmp2}
\end{equation}
by virtue of (\ref{twisted-r-14}) and (\ref{twisted-r-12}). Replacing $u$
and $v$ in (\ref{B-B:tmp2}) by $x_i$ and $x_j$ respectively which are not
dynamical, we obtain
\begin{equation}
    [\tilde B^\ast(x_i;\utilde), \tilde B^\ast(x_j;\utilde)]
    =
    \frac{\theta'_{11} \theta_{11}(x_j+\utilde)}
         {\theta_{11}(x_j) \theta_{11}(\utilde)} \tilde B^\ast(x_i)
    -
    \frac{\theta'_{11} \theta_{11}(x_i+\utilde)}
         {\theta_{11}(x_i) \theta_{11}(\utilde)} \tilde B^\ast(x_j),
\label{consistency:B-B}
\end{equation}
which means that the equation (\ref{B-Phi}) for $i$ and for $j$ are
compatible. 

Next we show the compatibility condition
\begin{equation}
    [X_i + \tilde A^\ast(x_i;\utilde),
     X_j + \tilde A^\ast(x_j;\utilde)] = 0,
\label{consistency:A-A}
\end{equation}
which implies the consistency of the equations (\ref{A-Phi}) for $i$
and for $j$ ($i\neq j$). It is obvious from (\ref{def:Xi}) that
\begin{equation}
    [X_i, X_j]=0.
\label{X-X}
\end{equation}
Because of (\ref{fix:utilde}), we have
$$
    [X_i, \tilde A^\ast(x_j;\utilde)]
    =
    - \left( \frac{\partial}{\partial \utilde} \tilde A (x_j;\utilde)
      \right)^\ast.
$$
By the same argument as that for (\ref{der-utilde-AT}) the right hand side
is rewritten as
\begin{equation}
    [X_i, \tilde A^\ast(x_j;\utilde)]
    =
    -\frac{\theta'_{11}\theta_{11}(\utilde + x_j)}
          {2\theta_{11}(\utilde)\theta_{11}(x_j)}
     \tilde C^\ast(x_j;\utilde)
    -\frac{\theta'_{11}\theta_{11}(\utilde - x_j)}
          {2\theta_{11}(\utilde)\theta_{11}(x_j)}
     \tilde B^\ast(x_j;\utilde).
\label{X-i-A-j}
\end{equation}
Exchanging $i$ and $j$, we have
\begin{equation}
    [X_j, \tilde A^\ast(x_i;\utilde)]
    =
    -\frac{\theta'_{11}\theta_{11}(\utilde + x_i)}
          {2\theta_{11}(\utilde)\theta_{11}(x_i)}
     \tilde C^\ast(x_i;\utilde)
    -\frac{\theta'_{11}\theta_{11}(\utilde - x_i)}
          {2\theta_{11}(\utilde)\theta_{11}(x_i)}
     \tilde B^\ast(x_i;\utilde).
\label{X-j-A-i}
\end{equation}
The (1,1)-element of (\ref{comm-rel:twisted-T}) means
\begin{eqnarray}
\hspace{-1cm}
[\tilde A^\ast(x_i;\utilde), \tilde A^\ast(x_j;\utilde)] &=&
    - \tilde r_{13}(x_i,x_j;\utilde) \tilde C^\ast(x_i;\utilde)
    - \tilde r_{12}(x_i,x_j;\utilde) \tilde C^\ast(x_j;\utilde)
\nonumber \\
    && + \tilde r_{31}(x_i,x_j;\utilde) \tilde B^\ast(x_i;\utilde)
    + \tilde r_{21}(x_i,x_j;\utilde) \tilde B^\ast(x_j;\utilde).
\label{A-A:tmp}
\end{eqnarray}
Summing up (\ref{X-X}), (\ref{X-i-A-j}),
(\ref{X-j-A-i}) and (\ref{A-A:tmp}), we have proved
(\ref{consistency:A-A}) because of (\ref{twisted-r-12}).

The consistency of (\ref{A-Phi}) for $i$ and (\ref{B-Phi}) for $j$ is
shown as follows. First assume $i\neq j$. Then the same computation as
above gives
\begin{eqnarray}
\lefteqn{
    [X_i+\tilde A^\ast(x_i;\utilde), \tilde B^\ast(x_j;\utilde)
\,=\,    - \left(
      \frac{\partial}{\partial \utilde} \tilde B(x_j;\utilde)
      \right)^\ast
    + [\tilde A^\ast(x_i;\utilde), \tilde B^\ast(x_j;\utilde)]
}    \nonumber \\
&=&
    \left(\frac{\theta'_{11}(x_i-x_j)}{\theta_{11}(x_i-x_j)}
         -\frac{\theta'_{11}(x_i    )}{\theta_{11}(x_i    )}\right)
    \tilde B^\ast(x_j;\utilde)
-\frac{\theta'_{11}\theta(x_i-x_j-\utilde)}
         {\theta_{11}(x_i-x_j)\theta_{11}(\utilde)}
    \tilde B^\ast(x_i;\utilde).
\label{consistency:A-i-B-j}
\end{eqnarray}
Thus we have proved the compatibility of (\ref{A-Phi}) for $i$ and
(\ref{B-Phi}) for $j$. Here we used
\begin{equation}
    \frac{\partial}{\partial \utilde} \tilde B(x_j;\utilde)
    =
    - \frac{\theta'_{11}\theta_{11}(x_j+\utilde)}
           {\theta_{11}(x_j)\theta_{11}(\utilde)}
      \tilde A(x_j;\utilde)
    + \frac{\theta'_{11}(x_j)}{\theta_{11}(x_j)}
      \tilde B(x_j;\utilde),
\label{der-utilde-B}
\end{equation}
and the (1,2)-element of (\ref{comm-rel:twisted-T}).

The case $i=j$ is almost the same, but there is another term coming from 
$[X_i, \tilde B^\ast(x_i;\utilde)]$:
\begin{eqnarray}
\lefteqn{
    [X_i + \tilde A^\ast(x_i;\utilde),
    \tilde B^\ast(x_i;\utilde)]
    =}\nonumber\\
    &=&
      \left.\frac{\partial}{\partial u}\right|_{u=x_i}\!\!\!
      \tilde B^\ast(u;\utilde)
    - \left(
      \frac{\partial}{\partial \utilde} \tilde B(x_i;\utilde)
      \right)^\ast
    + [\tilde A^\ast(x_i;\utilde), \tilde B^\ast(x_i;\utilde)].
\label{A-i-B-i:tmp}
\end{eqnarray}
By the same computation as (\ref{(A-C-TT)jj}), it follows from the
(1,2)-element of (\ref{comm-rel:twisted-T}) that
\begin{eqnarray}
\lefteqn{
    [\tilde A^\ast(x_i;\utilde), \tilde B^\ast(x_i;\utilde)]
    =}\nonumber\\
    &=&
    \frac{\theta'_{11}(\utilde)}{\theta_{11}(\utilde)} 
    \tilde B^\ast(x_i;\utilde)
    -
    \frac{\theta'_{11}\theta_{11}(x_i+\utilde)}
         {\theta_{11}(x_i)\theta_{11}(\utilde)}
    \tilde A^\ast(x_i;\utilde)
    -
    \left. \frac{\partial}{\partial u}\right|_{u=x_i}\!\!\!
    \tilde B^\ast(u;\utilde).
\label{[A-i,B-i]}
\end{eqnarray}
Substituting (\ref{[A-i,B-i]}) and (\ref{der-utilde-B}) for $j=i$ into
(\ref{A-i-B-i:tmp}), we obtain
\begin{equation}
    [X_i+\tilde A^\ast(x_i;\utilde), 
     \tilde B^\ast(x_i;\utilde)]
    =
    \left(
    \frac{\theta'_{11}(\utilde)}{\theta_{11}(\utilde)}
    -
    \frac{\theta'_{11}(x_i)}{\theta_{11}(x_i)}
    \right)
    \tilde B^\ast(x_i;\utilde),
\label{consistency:A-B}
\end{equation}
which proves the consistency of (\ref{A-Phi}) for $i$ and (\ref{B-Phi})
for $i$.
Q.E.D.

\subsection{Separating operator}

The separating integral operator $\sov$ is defined by (\ref{def:sov})
with the kernel function $\Phi(x|y)$ satisfying equations (\ref{B-Phi})
and (\ref{A-Phi}). 

\begin{prop}
\label{quantum-separated-equation}
(i) For any function $f$ of $(y_1,\dots,y_N)$ in $V_1\tensor \cdots
\tensor V_N$, we have
\begin{eqnarray}
    \sov( \tilde B(x_i;\tilde u) f) &=& 0,
\label{K(B(xi)f)=0}
\\
    \sov(-\tilde A(x_i;\tilde u) f) &=& X_i f.
\label{K(A(xi)f)=Xif}
\end{eqnarray}

(ii)
The elliptic Gaudin Hamiltonian $\hat\tau(u)$ with
the spectral parameter fixed to $u=x_i$ is transformed as follows.
\begin{equation}
    \sov(\hat\tau(x_i)f)(x) =
    X_j^2 \sov(f)(x),
\label{sov(tau(x-i)f)}
\end{equation}
where $X_i$ is defined by (\ref{def:Xi}).
\end{prop}

{\it Proof\/}.
(i) is a direct consequence of (\ref{B-Phi}) and (\ref{A-Phi})
respectively. 

(ii)
By the definition (\ref{def:tau})
\begin{eqnarray}
\lefteqn{    \sov(\hat\tau(x_i) f) (x) =  } \nonumber \\
&&    \frac12 \int \Phi(x|y)\,
    \bigl((2\tilde A(x_i;\tilde u)^2
          + \tilde B(x_i;\tilde u) \tilde C(x_i;\tilde u)
        + \tilde C(x_i;\tilde u) \tilde B(x_i;\tilde u))
    f(y)\bigr)
    \, dy
\nonumber\\
   &=& \int (\tilde A^\ast(x_i;\tilde u))^2 \Phi(x|y)\, f(y)\, dy
    + \int \tilde C^\ast(x_i;\tilde u)
             \tilde B^\ast(x_i;\tilde u) \Phi(x|y)\, f(y)\, dy
\nonumber\\
    &&+ \frac12
        \int [\tilde B^\ast(x_i;\tilde u), \tilde C^\ast(x_i;\tilde u)]
        \Phi(x|y)\, f(y)\, dy
\label{sov(tau(x-i)f):tmp}
\end{eqnarray}
The first term in the right hand side of (\ref{sov(tau(x-i)f):tmp}) is
rewritten by the following formula:
\begin{eqnarray}
    (\tilde A^\ast(x_i;\tilde u))^2 \Phi(x|y)
    &=&
    - \tilde A^\ast(x_i) X_i \Phi(x|y)
\nonumber\\
    &=&
    X_i^2 \Phi(x|y)
    +
    [X_i, \tilde A^\ast(x_i)] \Phi(x|y),
\label{A2Phi}
\end{eqnarray}
where we used (\ref{A-Phi}). The last term of (\ref{A2Phi}) is
\begin{equation}
    [X_i, \tilde A^\ast(x_i;\tilde u)]
    =
    \left. \frac{\partial}{\partial u}\right|_{u=x_i} 
    \tilde A^\ast(u;\tilde u)
    -
    \left. \frac{\partial}{\partial \tilde u}\right|_{u=x_i} 
    \tilde A^\ast(u;\tilde u)
\end{equation}
because $\tilde u = \sum z_n - \sum x_i$. Hence, similarly to the
derivation of (\ref{X-i-A-j}), we can prove that
\begin{eqnarray}
    [X_i, \tilde A^\ast(x_i;\tilde u)]
    &=&
    \left. \frac{\partial}{\partial u}\right|_{u=x_i} 
    \tilde A^\ast(u;\tilde u)
    -\frac{\theta'_{11}\theta_{11}(\tilde u+ x_i)}
          {2\theta_{11}(\tilde u)\theta_{11}(x_i)}
     \tilde C^\ast(x_i;\tilde u) \nonumber\\
    &-& \frac{\theta'_{11}\theta_{11}(\tilde u - x_i)}
          {2\theta_{11}(\tilde u)\theta_{11}(x_i)}
     \tilde B^\ast(x_i;\tilde u).
\label{X-i-A-i}
\end{eqnarray}

The $(2,3)$-element of the commutation relation (\ref{comm-rel:twisted-T})
gives
\begin{eqnarray}
    [\tilde B(u;\tilde u), \tilde C(u;\tilde u)]
    &=& 2 \tilde A'(u;\tilde u)
    -
    \frac{\theta'_{11} \theta_{11}(u-\tilde u)}
         {\theta_{11}(\tilde u) \theta_{11}(u)} \tilde B(u;\tilde u)
    \nonumber\\
    &-&
    \frac{\theta'_{11} \theta_{11}(u+\tilde u)}
         {\theta_{11}(\tilde u) \theta_{11}(u)} \tilde C(u;\tilde u).
\label{B-C}
\end{eqnarray}
in the limit $v \to u$. Substituting (\ref{A2Phi}), (\ref{X-i-A-i})
and (\ref{B-C}) into (\ref{sov(tau(x-i)f):tmp}) and using (\ref{B-Phi}),
we obtain (\ref{sov(tau(x-i)f)}).
Q.E.D.

\medskip
The equation (\ref{K(B(xi)f)=0}) is a quantum version of
(\ref{zero-of-tilde-B}) and the equation (\ref{K(A(xi)f)=Xif}) together
with the canonical commutation relation $[X_i, x_j] = \delta_{ij}$ means
that operators $(x_1, \dots, x_N; X_1, \dots, X_N)$ are quantization of
the classical separated variables in \secref{poisson-rel}.

The second statement of \propref{quantum-separated-equation} provides a
formal separation of variables for the quantum elliptic Gaudin model.
Using the language of \cite{fre:95} and \cite{e-f-r:98}, the kernel
$\Phi(x|y)$ provides a Radon-Penrose transformation of the corresponding
$D$-modules (cf.\ \cite{dag-sch:96}).

In principle, the quantum separation of variables should result in
a one dimensional spectral problem for the separated equation
(\ref{sov(tau(x-i)f)}) which is equivalent to the spectral problem for the
original Hamiltonians (\ref{def:H-n}). To achieve this goal one needs
to specify an integration contour in (\ref{def:sov}) to study in detail the
action of the integral operator $\sov$ on the functional space $V$.
Here we examine only the simplest case $N=1$, leaving the general case for
the further study.

\section{Quantum system: case $N=1$.}
\label{N=1}
\setcounter{equation}{0}

In this section we examine the
special case of $N=1$. In this case, everything can be computed explicitly
and we shall see that the separated equation is nothing but the classical
Lam\'e equation and its generalization.

We adopt the realization of the representation $\rho^\ell$ of $sl_2$ on
the space of elliptic functions reviewed in
\appref{rep-on-elliptic-curve}. We could use the standard realization on the
space of sections of a line bundle over $\Proj^1$, but the result is
essentially the same up to coordinate transformation and gauge
transformation. We omit the suffix $n$ of $z_n$ and $S^a_n$ for
brevity.

\subsection{Separated variables}
\label{SoV:N=1}

The quantum twisted $B$ operator
$\tilde B(u;\utilde) = \tilde\calB(u;\utilde)$ is defined as in the
classical case (\ref{def:tilde-T}) or (\ref{def:tilde-L}). Substituting
(\ref{def:L}) we obtain
\begin{eqnarray}
    \tilde B(u;\utilde) &=&
    \frac{\theta'_{11}}
    {2\theta_{11}(u)\theta_{11}(\utilde)\theta_{11}(u-z)}
    \Bigl(  \theta_{10}(u-z)\theta_{10}(u+\utilde) S^1 \nonumber\\
    &&-\theta_{00}(u-z)\theta_{00}(u+\utilde) iS^2
    -\theta_{01}(u-z)\theta_{01}(u+\utilde) S^3\Bigr).
\label{tilde-B:N=1}
\end{eqnarray}
The realization of the representation (\ref{def:S-y}) gives the following
expression: 
\begin{equation}
    \tilde B(u;\utilde) =
    \tilde B^{(1)}(u;\utilde) \frac{d}{dy} + \tilde B^{(0)}(u;\utilde),
\label{tilde-B:N=1:diff-op}
\end{equation}
where
\begin{eqnarray}
    \tilde B^{(1)}(u;\utilde)&=&
    \theta_{11}(u)^{-1} \theta_{11}(\utilde)^{-1}
    \theta_{11}(u-z)^{-1} \theta_{11}(2y)^{-1}
    \nonumber\\
&&\times
    \theta_{10}\left(y+u-\frac{z}{2}+\frac{\utilde}{2}\right)
    \theta_{10}\left(y-u+\frac{z}{2}-\frac{\utilde}{2}\right)
\nonumber \\
&&\times
    \theta_{10}\left(-y -\frac{z}{2}-\frac{\utilde}{2}\right)
    \theta_{10}\left(-y +\frac{z}{2}+\frac{\utilde}{2}\right)
\label{tilde-B:N=1:(1)}
\end{eqnarray}
\begin{eqnarray}
 \lefteqn{   \tilde B^{(0)}(u;\utilde)\,=\,
    \frac{2\ell\theta'_{11}}
    {2\theta_{11}(u)\theta_{11}(\utilde)\theta_{11}(u-z)\theta_{11}(y)^2}
\Biggl(\theta_{11}(u+\utilde)\theta_{11}(u-z)\theta_{11}(y)^2
    }  \nonumber\\
    && +2 \theta_{10}\left( y+u-\frac{z}{2}+\frac{\utilde}{2}\right)
       \theta_{10}\left(-y+u-\frac{z}{2}+\frac{\utilde}{2}\right)
       \theta_{10}\left(    -\frac{z}{2}-\frac{\utilde}{2}\right)^2
    \Biggr)
\label{tilde-B:N=1:(0)}
\end{eqnarray}

A special point in the case $N=1$ is that we can make use of the freedom
of $\utilde$ so that $\tilde B(u;\utilde)$ is a multiplication operator
with the divisor of the form,
\begin{equation}
    \div(\tilde B(u;\utilde)) =  [x] + [z] - [z] - [0] = [x]-[0]
                      \pmod{\Integer+\tau\Integer},
\label{tilde-B:div:N=1}
\end{equation}
as in the classical case, (\ref{def:x-j}), (\ref{sum-x-j}).  In fact, if
we put $\utilde = -z \pm 2y +1$, $\tilde B^{(1)}(u;\utilde)=0$ by virtue
of (\ref{tilde-B:N=1:(1)}), and then (\ref{tilde-B:N=1:(0)}) implies
\begin{equation}
    \tilde B(u;\utilde)|_{\utilde = -z\pm 2y+1} =
    \frac{2\ell\theta'_{11}\theta_{11}(u-z\pm 2y)}
         {-2\theta_{11}(u)\theta_{11}(-z\pm 2y)}.
\label{def:x-j:q:N=1}
\end{equation}
(We substitute the variable ``from the left'', namely we define
$$ 
    \tilde B(u;\utilde)|_{\utilde = -z\pm 2y+1} =
    \tilde B^{(1)}(u;-z\pm 2y+1) \frac{d}{dy} +
    \tilde B^{(0)}(u;-z\pm 2y+1).
$$ 
Hereafter we always follow this normal ordering convention.)
Therefore we can take $x=z\mp 2y$ in (\ref{tilde-B:div:N=1}). This is the
one of the ``separated variables'' in this case.

In the classical model, \thmref{canonical-poisson-relation},
$-\tilde A(x;\utilde)$ is a dynamical variable canonically conjugate to
$x$. This is also the case in the quantum model.  
The definition of $\tilde A$, (\ref{def:tilde-T}), is rewritten in the
form 
\begin{eqnarray}
 \lefteqn{   \tilde A(u;\utilde) \,=\,
    \frac{\theta'_{11}}
    {2\theta_{11}(u)\theta_{11}(\utilde)\theta_{11}(u-z)}
    \Bigl(
    \theta_{10}^{-1}\theta_{10}(u-z)\theta_{10}(u)\theta_{10}(\utilde)S^1
 }   \nonumber \\
&&-\theta_{00}^{-1}\theta_{00}(u-z)\theta_{00}(u)\theta_{00}(\utilde)iS^2
 -\theta_{01}^{-1}\theta_{01}(u-z)\theta_{01}(u)\theta_{01}(\utilde)S^3\Bigr)
\label{tilde-A:N=1}
\end{eqnarray}
by (\ref{def:L}). Substituting $\utilde = -z\pm 2y + 1$ and $u=x=z\mp 2y$
(from the left), we obtain
\begin{eqnarray}
\lefteqn{
    \tilde A(u;\utilde)|_{u=z\mp 2y, \utilde = -z\pm 2y+1}
    =} \nonumber \\ &\hspace{1.5cm}=&
    \pm \frac{1}{2}
    \left(
    \frac{d}{dy} +
    2\ell\left(
     \frac{\theta'_{11}\theta_{10}(2y)}{\theta_{10}\theta_{11}(2y)} +
     \frac{\theta'_{11}\theta_{00}(2y)}{\theta_{00}\theta_{11}(2y)} +
     \frac{\theta'_{11}\theta_{01}(2y)}{\theta_{01}\theta_{11}(2y)} 
     \right)
    \right)
    \nonumber\\
    &\hspace{2cm}=&
    \pm \frac{1}{2}
    \left( \frac{d}{dy} - \ell \frac{\wp''(y)}{\wp'(y)} \right).
\label{tilde-A(x):q:N=1}
\end{eqnarray}
The last equality can be proved by comparing the poles of both hand
sides. Therefore $x = z\mp 2y$ and $X:=-\tilde A(u;\utilde)|_{u=z\mp 2y,
\utilde = -z\pm 2y+1}$ are the canonical conjugate variables
satisfying,
\begin{equation}
    [X ,\; x] = 1.
\label{ccr:N=1}
\end{equation}
We did not make use of the formulation in the previous sections
explicitly. In fact, thanks to the special choice of $\utilde$, $\Phi$ in
(\ref{def:sov}) is a $\delta$-function type kernel, which reduces the
integral operator $\sov$ to a coordinate transformation operator from $y$
to $x$.

\subsection{Solving the spectral problem}

For the case $N=1$, the generating function of the quantum integrals of
motion $\hat\tau(u)$ ($u$ is the spectral parameter)
\begin{equation}
    \hat \tau(u) = \frac12 \sum_{a=1}^3 w_a(u)^2 (\rho^{(\ell)}(S^a))^2
\label{tau:N=1}
\end{equation}
is explicitly written down. Here we shift the spectral parameter in the
original definition (\ref{def:tau}) as $u \mapsto u_z$ and set $z=0$ for
the sake of simplicity. Using (\ref{def:S-y}) or (\ref{def:S:eta}) and
various identities of elliptic functions in \cite{whit-wat}, we can expand
the right hand side of (\ref{tau:N=1}):
\begin{equation}
    \hat\tau\left(y,\frac{d}{dy},\frac{d^2}{dy^2}; u \right)
    =\frac{1}{4}
    \left(
    \frac{d^2}{dy^2} 
    - 2\ell \frac{\wp''(y)}{\wp'(y)} \frac{d}{dy}
    + 4\ell(2\ell - 1) \wp(y) + 4 \ell (\ell + 1) \wp(u)
    \right)
\label{tau:y:N=1}
\end{equation}
or
\begin{eqnarray}
 \hspace{-1cm}
    \hat\tau\left(\eta,\frac{d}{d\eta},\frac{d^2}{d\eta^2}; \lambda \right)
    &=&
    (\eta - e_1)(\eta - e_2)(\eta - e_3)
    \nonumber\\
    &\times &
    \left(
    \frac{d^2}{d\eta^2}
    + \frac{1-2\ell}{2}
      \left(
      \frac{1}{\eta - e_1} + \frac{1}{\eta - e_2} + \frac{1}{\eta - e_3}
      \right)
      \frac{d}{d\eta}+ \right. \nonumber\\
    &&\left.
    + \frac{\ell (2\ell - 1) \eta + \ell (\ell + 1) \lambda}
           {(\eta - e_1)(\eta - e_2)(\eta - e_3)}
    \right),
\label{tau:eta:N=1}
\end{eqnarray}
where $\lambda = \wp(u)$.

As is expected from \propref{quantum-separated-equation} and the result for
the rational Gaudin model in \cite{skl:87}, operator $\hat \tau(u)$ is
factorized as follows when the spectral parameter $u$ is fixed to a
separated variable $x_1=2y$. (We may also take $x_1=-2y$.):
\begin{equation}
    \left.
    \hat\tau\left(y,\frac{d}{dy},\frac{d^2}{dy^2}; u \right)
    \right|_{u=2y}
    = \bigl(-\tilde A(u;\tilde u)|_{u=2y, \tilde u=2y+1}
      \bigr)^2
    = X^2,
\label{factor-tau:N=1}
\end{equation}
which immediately follows from (\ref{tilde-A:N=1}). (The operator $X$ is
defined before (\ref{ccr:N=1}).) This is consistent with the general
result (\ref{sov(tau(x-i)f)}).

Equations (\ref{tau:y:N=1}) or (\ref{tau:eta:N=1}) shows that the spectral
problem of the elliptic Gaudin model with $N=1$ is an ordinary
differential equation of second order on the elliptic curve
$\Complex/\Integer+\tau\Integer$:
\begin{equation}
    \hat\tau\left(y,\frac{d}{dy},\frac{d^2}{dy^2}; u \right) \psi(y)
    =
    t(u) \psi(y),
\label{spec-prob:N=1:y}
\end{equation}
or on the projective line $\Proj^1 (\Complex)$:
\begin{equation}
    \hat\tau\left(\eta,\frac{d}{d\eta},\frac{d^2}{d\eta^2}; \lambda \right)
    \psi(\eta)
    =
    t(\lambda) \psi(\eta).
\label{spec-prob:N=1:eta}
\end{equation}
Here $t(u)$, $t(\lambda)$ are eigenvalues of $\hat\tau$,
$\psi\in V^{(\ell)}$ is an eigenvalue corresponding to this eigenvalue.
Since operators $\hat\tau(u)$ and $U_\a$ commute with each other
by virtue of (\ref{def:X-a}, \ref{X=Ad-U}) and (\ref{tau:N=1}),
we can decompose each eigenspace of $\hat\tau(u)$ into those of $U_\a$.

The equation (\ref{spec-prob:N=1:y}) has regular singularities: $u=0$
$\pmod\Gamma$ with exponents $-4\ell$, $-2\ell+1$, and $u=\omega_\a$
($\a=1,2,3$, $\omega_1=1$, $\omega_2=\tau$, $\omega_3=1+\tau$) with
exponents $0$, $2\ell+1$.  The equation (\ref{spec-prob:N=1:eta}) has
regular singularities: $\eta = e_\a$ with exponents $0$, $(2\ell+1)/2$, and
$\eta = \infty$ with $\half - \ell$, $-2\ell$.

If $\ell$ is an integer, these equations are ordinary Lam\'e equations,
while for $\ell \in \half+\Integer$ they are generalized Lam\'e equations
studied by Brioschi, Halphen and Crawford.
Following the classical theory of Lam\'e functions
(see Chap.~XXIII \cite{whit-wat}), we can solve the spectral problem
(\ref{spec-prob:N=1:y}), (\ref{spec-prob:N=1:eta}) in $V^{(\ell)}$
as follows.

\subsubsection{Case $\ell\in\Integer$.}

We want a solution $\psi(\eta)$ of (\ref{spec-prob:N=1:eta}) such that
$\psi(\eta) \in V^{(\ell)}$. Let us assume that $\psi(\eta)$ is expanded
around the singular point $e_\a$ as
\begin{equation}
    \psi(\eta) = \sum_{r = 0}^\infty a^\a_r (\eta - e_\a)^{2\ell - r},
\label{expansion:psi}
\end{equation}
$a_0$ being $1$.
The condition $\psi(\eta) \in V^{(\ell)}$ means that $a^\a_r = 0$
for $r > 2\ell$. Substituting (\ref{expansion:psi}) 
into (\ref{spec-prob:N=1:eta}), we obtain the following recursion
relation
\begin{eqnarray}
 \hspace{-0.8cm}
    r ( \ell + \texthalf - r ) a^\a_r
    &=&
    \Bigl( \bigl(\ell(2\ell-1) - 3(r-1)(2\ell-r+1)\bigr) e_\a + E
    \Bigr) a^\a_{r-1}
    \nonumber\\
    &&+
    (2\ell-r+2) \left(\ell-r+\frac{3}{2} \right) 
    (e_\a - e_\b) (e_\a - e_\g) a^\a_{r-2}
\label{recursion:ell=integer}
\end{eqnarray}
for $r>0$ where $E= \ell(\ell+1) \lambda - t(\lambda)$.
(Undefined coefficients $a^\a_r$ for $r<0$ are $0$.)
Hence, as a function of $E$, $a^\a_r= a^\a_r(E)$ is a polynomial of degree
$r$ of the form
\begin{equation}
    a^\a_r(E) = A_r E^r + O(E^{r-1}), \qquad
    A_r = \left(r!\prod_{j=1}^r\left(\ell-r-\half+j\right)\right)^{-1}
\label{def:A-r}
\end{equation}
Let us denote the roots of $a^\a_{2\ell+1}(E) = 0$ by $E^\a_i$
($i=1, \dots, 2\ell+1$). The recursion relation
(\ref{recursion:ell=integer}) implies $a^\a_r(E^\a_i)=0$ for $r\geq 2\ell+1$.
Hence we obtain a polynomial solution $\psi(\eta) = \psi(\eta;E^\a_i)$ of
(\ref{spec-prob:N=1:eta}) of the form (\ref{expansion:psi}) for each
$i=1,\dots,2\ell+1$, provided that
\begin{equation}
    t(\lambda) = \ell(\ell+1)\lambda - E^\a_i.
\label{spec:ell=integer}
\end{equation}

Conversely, if $\psi(\eta) \in V^{(\ell)}$ is a solution of the spectral
problem (\ref{spec-prob:N=1:eta}), then there exists certain $i$ for each
$\a=1,2,3$ such that $\psi(\eta) = \psi(\eta;E^\a_i)$.
This is proved by expanding the polynomial $\psi(\eta)$
as in (\ref{expansion:psi}) and tracing back the above argument.

\begin{prop}
\label{non-deg:ell=integer}
Assume that $\omega_2 = \tau$ is pure imaginary and that parameters $z_n$
are all real numbers. Then all $E^\a_i$ are real and the spectral problem
(\ref{spec-prob:N=1:eta}) is non-degenerate.  Namely $E^\a_i \neq E^\a_j$
for distinct $i$, $j$ and the solutions $\psi(\eta;E^\a_i)$ span the space
$V^{(\ell)}$.  In particular $E^\a_i$ ($i=1,\dots,2\ell+1$) for $\a=1,2,3$
coincide up to order, and
$a^1_{2\ell+1}(E)=a^2_{2\ell+1}(E)=a^3_{2\ell+1}(E)$.  Hence we can omit
the index $\a$ for $E^\a_i$ and $a^\a_{2\ell+1}(E)$.

Vector $\psi(\eta;E_i)$ is an eigenvector of $U_\a$ with eigenvalue
$(-1)^{\ell}$ if $a^\a_{\ell}(E_i)\neq0$ and
$(-1)^{\ell+1}$ if $a^\a_{\ell}(E_i) = 0$.
\end{prop}

{\it Proof\/}.
Under the assumption $\tau \in i\Real$, operator $\hat\tau(u)$
($u\in\Real$) is an hermitian operator because of (\ref{self-adjoint}),
and hence it is obvious that $E^\a_i$ are real and that
$\psi(\eta;E^\a_i)$ span $V^{(\ell)}$.

In order to show non-degeneracy of the spectral problem
(\ref{spec-prob:N=1:eta}) we have only to prove that $E^2_i$ are
distinct with each other. Define
$$
    {\tilde a}^2_r (E) :=
    \left\{
    \begin{array}{rl}
     a^2_r(E),         \quad&             r <     \ell+1, \\
    (-1)^{r-l}a^2_r(E),\quad& \ell+1 \leq r \leq 2\ell+1.
    \end{array}\right.
$$
Then the leading coefficients of ${\tilde a}^2_r$ is
\begin{equation}
    {\tilde a}^2_r(E) = {\tilde A}_r E^r + O(E^{r-1}), \qquad
    {\tilde A}_r = |A_r|.
\label{def:tilde-A-r}
\end{equation}
The recursion relation (\ref{recursion:ell=integer}) is rewritten as
\begin{equation}
    c_r {\tilde a}^2_r(E) =
    q_r {\tilde a}^2_{r-1}(E) - k_{r-2} {\tilde a}^2_{r-2}(E),
\label{recursion:ell=integer:2}
\end{equation}
where
\begin{eqnarray}
    c_r &=& r |\ell + \texthalf - r|, 
\nonumber\\
    q_r &=& \bigl(\ell(2\ell-1) - 3(r-1)(2\ell-r+1)\bigr) e_\a + E,
\label{def:coef-of-recursion}\\
    k_r &=& \left| \ell - r + \frac{3}{2} \right| (2\ell - r + 2)
          (e_1 - e_2) (e_2 - e_3).
\nonumber
\end{eqnarray}
Since $e_1 > e_2 > e_3$ under the assumption of the proposition, we have
$c_r > 0$ and $k_r > 0$. This fact together with $\tilde A_r > 0$ (see
(\ref{def:tilde-A-r})) implies that all the roots of ${\tilde a}^2_r(E)$
are real and distinct by Sturm's theorem (see, e.g., Chap.~IX, \S\S 4--5,
\cite{dick}). This proves the first statement of the proposition.

The operators $U_\a$ and $\hat\tau$ commute and each eigenspace of
$\hat\tau$ is one-dimensional. Hence $\psi(\eta;E_i)$ is an eigenvector
of $U_\a$. Recall that $U_\a$ has eigenvalues $(-1)^\ell$
with multiplicity $\ell+1$ and $(-1)^{\ell+1}$ with multiplicity
$\ell$. (See \S\ref{involution}.) If $a^\a_\ell(E_i) \neq 0$, then
$$
    U_\a \psi(\eta;E_i) = (-1)^\ell \psi(\eta;E_i)
$$
because of (\ref{U-on-basis}). Hence there are at most $\ell+1$ of $E_i$'s
such that $a^\a_\ell(E_i) \neq 0$. In other words, at least $\ell$ of
$E_i$'s satisfy $a^\a_\ell(E_i)=0$. Since $a^\a_\ell(E)$ is 
a polynomial of degree $\ell$, this proves the second statement of the
proposition.
Q.E.D.

\subsubsection{Case $\ell\in \half + \Integer$.}

As in the case $\ell\in\Integer$, we consider an expansion
(\ref{expansion:psi}) of a solution $\psi(\eta)$ of the spectral problem
(\ref{spec-prob:N=1:eta}), but this time we consider the series which
terminate at $r=\ell - \half$:
\begin{equation}
    \psi(\eta) = \sum_{r = 0}^{\ell-1/2} a^\a_r (\eta - e_\a)^{2\ell - r},
\label{expansion:psi:ell=half-odd}
\end{equation}
They are parametrized by zeros of
the polynomial $a^\a_{\ell+\half}(E)$, $\{E^\a_i\}_{i=1, \dots, \ell+\half}$
as in the previous case: $\psi(\eta) = \psi(\eta;E^\a_i)$.

Another set of solutions are obtained from this set by applying the operator
$U_\a$:
\begin{equation}
    U_\a \psi(\eta;E^\a_i)
    = \sum_{r=0}^{\ell - \half} {a^\a}'_r(E^\a_i) (\eta - e_\a)^r,
\label{expansion:U-psi}
\end{equation}
since $U_\a$ and $\hat\tau(u)$ commute.

The following proposition is proved in the same manner as
\propref{non-deg:ell=integer}.

\begin{prop}
Assume that $\omega_2 = \tau$ is pure imaginary and that parameters $z_n$
are all real numbers.

Then all $E^\a_i$ are real and $E^\a_i \neq E^\a_j$ for distinct $i$, $j$.

The solutions $\psi(\eta;E^\a_i)$ and $U_\a\psi(\eta;E^\a_i)$ span
the space $V^{(\ell)}$.
In particular $E^\a_i$ ($i=1,\dots,\ell+\half$) for $\a=1,2,3$ coincide
up to order, and
$a^1_{\ell+\half}(E)=a^2_{\ell+\half}(E)=a^3_{\ell+\half}(E)$.
Hence we can omit the index $\a$ for $E^\a_i$ and $a^\a_{\ell+\half}(E)$.

Vectors $\psi(\eta;E_i) \pm U_\a \psi(\eta;E_i)$ are eigenvectors
of $U_\a$ with eigenvalues $\mp i$.
\end{prop}

This proposition means that each eigenvalue $E_i$ degenerates with
multiplicity two. It was Crawford \cite{craw} who first found the relation
of these two solutions (one is obtained from the other by operating $U_2$)
by the explicit expansions of type (\ref{expansion:psi:ell=half-odd}),
(\ref{expansion:U-psi}). See also p.578 of \cite{whit-wat}.

\appendix

\section{Notations}
\label{notations}
\setcounter{equation}{0}

We use the notation for the theta functions with characteristics as
follows (see \cite{mum}): for $a,b = 0,1$,
\begin{equation}
    \theta_{ab}(u;\tau)
    =
    \sum_{n\in\Integer} 
    e^{\pi i (n+a/2)^2 \tau + 2 \pi i (n+a/2)(u+b/2)}.
\label{def:theta}
\end{equation}
Unless otherwise specified, $\theta_{ab}(u) = \theta_{ab}(u;\tau)$. We 
also use abbreviations
\begin{equation}
    \theta_{ab} = \theta_{ab}(0), \qquad
    \theta_{ab}'= \left.\frac{d}{du}\right|_{u=0} \theta_{ab}(u).
\label{def:theta-zero}
\end{equation}
Quasi-periodicity properties of theta functions:
\begin{equation}
  \theta_{ab}(u) = (-1)^a \theta_{ab}(u+1) 
                 = e^{\pi i \tau + 2\pi i u} \theta_{ab}(u+\tau).
\label{period:theta}
\end{equation}
Parity of thetas:
$$ \theta_{00}(-u)=\theta_{00}(u), \quad
   \theta_{01}(-u)=\theta_{01}(u), \quad
   \theta_{10}(-u)=\theta_{10}(u), \quad
   \theta_{11}(-u)=-\theta_{11}(u)
$$

\subsection{Weierstrass functions}
Below we fix $\omega_1 = 1$ and $\omega_2 = \tau$.
\begin{equation}
    \s(u) = u\prod_{m,n\neq0}\left(1-\frac{u}{\omega_{mn}}\right)
               \exp\left[\frac{u}{\omega_{mn}}
                        +\half\left(\frac{u}{\omega_{mn}}\right)^2
                   \right]
\end{equation}
where $\omega_{mn}=m\omega_1+n\omega_2$.
$$
 \zeta(u)=\frac{\s^\prime(u)}{\s(u)}, \qquad \wp(u)=-\zeta^\prime(u).
$$
\begin{eqnarray*}
 \s(u+\omega_l)&=&-\s(u)e^{\eta_l(2u+\omega_l)} \\
 \zeta(u+\omega_l)&=&\zeta(u)+2\eta_l, \\
 \wp(u+\omega_l)&=&\wp(u).
\end{eqnarray*}
where
$
  \eta_l=\zeta\left(\omega_l/2\right),
$
which satisfy
$$
   \eta_1\omega_2-\eta_2\omega_1=\pi i.
$$

Sigma function is expressed by theta functions as follows:
$$
  \s(u) = \omega_1 e^{\eta_1 u^2/\omega_1}
          \frac{\theta_{11}(u/\omega_1)}{\theta_{11}'}.
$$

$$
 \s(-u)=-\s(u), \qquad \zeta(-u)=-\zeta(u), \qquad \wp(-z)=\wp(u).
$$
$$
 u\sim0: \qquad
 \s(u)=u+O(u^5), \qquad \zeta(u)=u^{-1}+O(u^3), \qquad
               \wp(u)=u^{-2}+O(u^2).
$$

Other sigma functions are defined as follows:
\begin{eqnarray*}
    \s_{00}(u) &=& e^{-(\eta_1+\eta_2)u}
    \frac{\s\left(u+\frac{\displaystyle\omega_1+\omega_2}
                         {\displaystyle 2}\right)}
         {\s\left(  \frac{\displaystyle\omega_1+\omega_2}
                         {\displaystyle2}\right)}
    = e^{\frac{\eta_1}{\omega_1}u^2}
      \frac{\theta_{00}(u/\omega_1)}{\theta_{00}(0)},
\\
    \s_{10}(u) &=& e^{-\eta_1u}
    \frac{\s\left(u+\frac{\displaystyle\omega_1}
                         {\displaystyle 2}\right)}
         {\s\left(  \frac{\displaystyle\omega_1}
                         {\displaystyle 2}\right)}
    = e^{\frac{\eta_1}{\omega_1}u^2}
      \frac{\theta_{10}(u/\omega_1)}{\theta_{10}(0)},
\\
    \s_{01}(u) &=& e^{-\eta_2u}
    \frac{\s\left(u+\frac{\displaystyle\omega_2}
                         {\displaystyle 2}\right)}
         {\s\left(  \frac{\displaystyle\omega_2}
                         {\displaystyle 2}\right)}
    = e^{\frac{\eta_1}{\omega_1}u^2}
      \frac{\theta_{01}(u/\omega_1)}{\theta_{01}(0)},
\end{eqnarray*}
which satisfy
$$
  \s_{g_1g_2}(u+\omega_l)=(-1)^{g_l}e^{\eta_l(2u+\omega_l)}\s_{g_1g_2}(u).
$$
$$
  \s_{g_1g_2}(-u)=\s_{g_1g_2}(u), \qquad \s_{g_1g_2}(0)=1
$$

Defining $e_1 = \wp(\omega_1/2)$, $e_2 = \wp((\omega_1+\omega_2)/2)$,
$e_3 = \wp(\omega_2/2)$, we have
$$ \frac{\s^2_{10}(u)}{\s^2(u)}+e_1=
   \frac{\s^2_{00}(u)}{\s^2(u)}+e_2=
   \frac{\s^2_{01}(u)}{\s^2(u)}+e_3=\wp(u),
$$
$$ e_1+e_2+e_3=0. $$
$$
    e_1 - e_2 =
    \left( \frac{\pi}{\omega_1}\right)^2 \theta_{01}(0)^4,
\quad
    e_1 - e_3 =
    \left( \frac{\pi}{\omega_1}\right)^2 \theta_{00}(0)^4,
\quad
    e_2 - e_3 =
    \left( \frac{\pi}{\omega_1}\right)^2 \theta_{10}(0)^4.
$$

We also use normalized Weierstra{\ss} functions:
\begin{equation}
    \zeta_{11}(u) = \frac{d}{du} \theta_{11}(u), \qquad
    \wp_{11}(u) = - \frac{d}{du} \zeta_{11}(u).
\label{def:zeta-p-11}
\end{equation}

\section{Realization of spin $\ell$ representations on an elliptic
curve.}
\label{rep-on-elliptic-curve}
\setcounter{equation}{0}

We recall here the following realization of the spin $\ell$ representation
of the Lie algebra $sl_2(\Complex)$. Let $e$, $f$, $h$ be the Chevalley
generators and define $S^1 = e+f$, $S^2 = -ie+if$ and $S^3 = h$. They
satisfy the relation $[S^a, S^b] = 2i S^c$ for any cyclic permutation
$(a,b,c)$ of $(1,2,3)$ and represented by the Pauli matrices $\sigma^a$. 

\subsection{Spin $\ell$ representations.}
\label{spin-l-rep-on-elliptic-curve}
The representation space $V^{(\ell)}$ is realized by
\begin{eqnarray}
\lefteqn{
    V^{(\ell)} = \bigoplus_{k=0}^{2\ell} \Complex \wp(y)^k
    }\nonumber\\
    &=&
    \{ \mbox{ even elliptic function } f(y) \,|\,
       \div(f) \geq -4\ell (\Integer+\tau\Integer) \}.
\label{def:V-l}
\end{eqnarray}
The generators $S^a$ act on this space as differential operators of first
order:
\begin{eqnarray}
    \rho^{(\ell)}(S^1) &=&
    \frac{\theta_{10}\theta_{10}(2y)}{\theta'_{11}\theta_{11}(2y)}
    \frac{d}{dy}
    + 2\ell
    \frac{\theta_{10}(y)^2}{\theta_{11}(y)^2}
\nonumber\\
    \frac{1}{i}\rho^{(\ell)}(S^2) &=&
    \frac{\theta_{00}\theta_{00}(2y)}{\theta'_{11}\theta_{11}(2y)}
    \frac{d}{dy}
    + 2\ell
    \frac{\theta_{00}(y)^2}{\theta_{11}(y)^2}
\nonumber\\
    \rho^{(\ell)}(S^3) &=&
    \frac{\theta_{01}\theta_{01}(2y)}{\theta'_{11}\theta_{11}(2y)}
    \frac{d}{dy}
    + 2\ell
    \frac{\theta_{01}(y)^2}{\theta_{11}(y)^2},
\label{def:S-y}
\end{eqnarray}
or in terms of usual Weierstra{\ss} functions,
\begin{eqnarray}
    \rho^{(\ell)}(S^1) &=& a_1
    \left(
    \frac{\s_{10}(2y)}{\s(2y)}\frac{d}{dy} + 2\ell (\wp(y)-e_1)
    \right),
\nonumber\\
    \rho^{(\ell)}(S^2) &=& a_2
    \left(
    \frac{\s_{00}(2y)}{\s(2y)}\frac{d}{dy} + 2\ell (\wp(y)-e_2)
    \right),
\nonumber\\
    \rho^{(\ell)}(S^3) &=& a_3
    \left(
    \frac{\s_{01}(2y)}{\s(2y)}\frac{d}{dy} + 2\ell (\wp(y)-e_3)
    \right),
\label{def:S:sigma}
\end{eqnarray}
where $e_a = \wp(\omega_{\bar a}/2)$ ($\bar a = 1,3,2$, $\omega_1 = 1$,
$\omega_2 = \tau$, $\omega_3 = 1+\tau$) for $a=1,2,3$ respectively and 
\begin{equation}
    a_1 = \frac{1}{\sqrt{e_1-e_2}\sqrt{e_1-e_3}},\quad
    a_2 = \frac{i}{\sqrt{e_1-e_2}\sqrt{e_2-e_3}},\quad
    a_3 = \frac{1}{\sqrt{e_2-e_3}\sqrt{e_1-e_3}}.
\label{def:a-i}
\end{equation}

This realization is equivalent to the realization on the space of
polynomials of degree $\leq 2\ell$ (or, sections of a line bundle on
$\Proj^1(\Complex)$),
$$ 
    e = x^2 \frac{d}{dx} - 2\ell x, \qquad
    f = - \frac{d}{dx}, \qquad
    h = 2x \frac{d}{dx} - 2\ell x,
$$ 
via a coordinate transformation,
$x = - \theta_{01}(y;\tau/2)/\theta_{00}(y;\tau/2)$, and a gauge
transformation: 
$$ 
    \{\mbox{polynomials in $x$}\} \owns
    \phi(x) \mapsto 
    \left(\frac{\theta_{00}(y;\tau/2)}{\theta_{11}(y;\tau)^2}\right)^n
    \phi(x(y)) \in V^{(\ell)}.
$$ 
Note that this is also obtained by a gauge transformation from a
quasi-classical limit of the representation of the Sklyanin algebra on
theta functions \cite{skl:83}.

The following expression is obtained from the coordinate
transformation $\eta = \wp(y)$:
\begin{equation}
    V^{(\ell)} = \bigoplus_{k=0}^{2\ell} \Complex \eta^k,
\label{def:V-l2}
\end{equation}
and $S^\a$ acts on $V^{(\ell)}$ as
\begin{equation}
    \rho^{(\ell)}(S^\a) = a_\a
    \left(
    ((e_\a-e_\b)(e_\a-e_\g)-(\eta-e_\a)^2)\frac{d}{d\eta}
    +
    2\ell (\eta-e_\a)
    \right),
\label{def:S:eta}
\end{equation}

Let us assume that $\tau$ is a pure imaginary number. Then, as is well
known (see, e.g., \cite{whit-wat}), $e_a$ are real numbers and 
$e_1 > e_2 > e_3$. This implies that $a_1$ and $a_3$ are real, while $a_2$
is purely imaginary.

We introduce the following hermitian form in this representation
space: for elliptic functions $f(y)$, $g(y)$ belonging
to $V^{(\ell)}$ defined by (\ref{def:V-l}), we define
\begin{equation}
    \langle f, g \rangle :=
    \int_C
    \overline{f(\bar y_2)} \, g(y_1) \, \mu(y_1,y_2)
\label{def:herm.form:y}
\end{equation}
where the 2-cycle $C$ is defined by
$$
    C:= \{(y_1,y_2)\in(\Complex/\Gamma)^2, y_2 = \bar y_1 \},
$$
and the 2-form $\mu(y_1,y_2)$ is defined by
\begin{eqnarray}
 \hspace{-1cm}
    \mu(y_1,y_2) &:=&
    (e_1-e_2)^{2(\ell+1)} (e_2-e_3)^{2(\ell+1)}
\nonumber \\
 &&\times
    \frac{\s(2y_2) \s(y_2)^{4\ell} \s(2y_1) \s(y_1)^{4\ell}}
         {\s_{00}(y_2-y_1)^{2(\ell+1)} \s_{00}(y_2+y_1)^{2(\ell+1)}}
    \frac{dy_2 \wedge dy_1}{4i}
\nonumber\\
    &=&
    \left( 1+ \frac{(\wp(y_2)-e_2)(\wp(y_1)-e_2)}
                       {(e_1-e_2)(e_2-e_3)}
    \right)^{-2(\ell+1)}
    \frac{\wp'(y_2)\wp'(y_1) dy_2 \wedge dy_1}{4i}.
\label{def:mu:y}
\end{eqnarray}
This is nothing but a twisted version of the inner product introduced
in \cite{skl:83}.
If we take the description of $V^{(\ell)}$ of the form (\ref{def:V-l2}),
this hermitian form is expressed as follows: 
\begin{equation}
    \langle f, g \rangle :=
    \int_{\Complex}
    \overline{f(\bar \eta)} \, g(\eta) \, \mu(\eta, \bar\eta)
\label{def:herm.form:eta}
\end{equation}
where the 2-form $\mu(\eta,\bar\eta)$ is defined by
$$
    \mu(\eta, \bar\eta) := 
    \left( 1+ \frac{(\bar\eta-e_2)(\eta-e_2)}
                       {(e_1-e_2)(e_2-e_3)}
    \right)^{-2(\ell+1)}
    \frac{d\bar\eta \wedge d\eta}{2i}.
$$
An orthogonal basis with respect to this inner product is given by
$\{ (\eta - e_2)^j \}_{j=0, \dots, 2\ell}$:
\begin{equation}
    \langle (\eta-e_2)^j, (\eta-e_2)^k \rangle
    =
    2\pi \frac{(2j)!! (4\ell-2j)!!}{(4\ell+2)!!} 
    (e_1-e_2)^{j+1} (e_2-e_3)^{j+1}
    \delta_{jk}.
\label{norm-of-basis}
\end{equation}
The generators $S^a$ of the Lie algebra $sl_2$ act on the space
$V^{(\ell)}$ as self-adjoint operators:
\begin{equation}
    \langle \rho^{(\ell)}(S^a) f, g\rangle
    =
    \langle f, \rho^{(\ell)}(S^a) g\rangle.
\label{self-adjoint}
\end{equation}
This was first proved in \cite{skl:83}, but we can check it directly
by using formula (\ref{norm-of-basis}).

Hence, if $u$ and $z_n$ are real numbers, the operator $\hat\tau(u)$
defined by (\ref{def:tau}) and the integrals of motion $H_n$ defined by
(\ref{def:H-n}) are hermitian operators on the Hilbert space $V$ with
respect to $\langle \cdot, \cdot \rangle$.

\subsection{Involutions.}
\label{involution}

There are involutive automorphisms of the Lie algebra $sl_2$ defined by
\begin{equation}
    X_a(S^b) = (-1)^{1-\delta_{ab}} S^b.
\label{def:X-a}
\end{equation}
These automorphisms are induced on the spin $\ell$ representations
as
\begin{equation}
    X_a (S^b) = U_a^{-1} S^b U_a,
\label{X=Ad-U}
\end{equation}
where operators $U_a: V^{(\ell)} \to V^{(\ell)}$ are
defined by
\begin{eqnarray}
    (U_1 f)(y) &=&
    e^{\pi i \ell} 
    \left(\frac{\wp(y)-e_1}{\sqrt{e_1-e_2}\sqrt{e_1-e_3}}\right)^{2\ell}
    f \left(y+ \frac{\omega_1}{2} \right),
\nonumber\\
    (U_2 f)(y) &=&
    e^{2 \pi i \ell} 
    \left(\frac{\wp(y)-e_2}{\sqrt{e_1-e_2}\sqrt{e_2-e_3}}\right)^{2\ell}
    f \left(y+ \frac{\omega_1 + \omega_2}{2} \right),
\label{def:U}\\
    (U_3 f)(y) &=&
    e^{- \pi i \ell} 
    \left(\frac{\wp(y)-e_3}{\sqrt{e_1-e_3}\sqrt{e_2-e_3}}\right)^{2\ell}
    f \left(y+ \frac{\omega_2}{2} \right),
\nonumber
\end{eqnarray}
for a elliptic function $f(y) \in V^{(\ell)}$. (cf.~\cite{skl:83}.)
They satisfy commutation relations
$$
    U_\a^2 = (-1)^{2\ell}, \qquad
    U_\a U_\b = (-1)^{2\ell} U_\b U_\a = U_\g
$$
for any cyclic permutation $(\a, \b, \g)$ of $(1,2,3)$.
The action of these operators on the bases
$\{ (\eta-e_\a)^j \}_{j=0, \dots, 2\ell}$ is:
\begin{eqnarray}
    U_1 (\eta - e_1)^j &=&
    e^{\pi i \ell} (e_1-e_2)^{j-\ell} (e_1-e_3)^{j-\ell}
    (\eta - e_1)^{2\ell-j},
\nonumber\\
    U_2 (\eta - e_2)^j &=&
    e^{\pi i (2\ell-j)} (e_1-e_2)^{j-\ell} (e_2-e_3)^{j-\ell}
    (\eta - e_2)^{2\ell-j},
\label{U-on-basis}\\
    U_3 (\eta - e_3)^j &=&
    e^{-\pi i \ell} (e_1-e_3)^{j-\ell} (e_2-e_3)^{j-\ell}
    (\eta - e_3)^{2\ell-j}.
\nonumber
\end{eqnarray}
Hence eigenvalues of $U_a$ are $(-1)^{\ell}$ with multiplicity $\ell+1$
and $(-1)^{\ell+1}$ with multiplicity $\ell$ if $\ell$ is an integer, and
$\pm i$ both with multiplicity $\ell+\half$ if $\ell$ is a half of an odd
integer.

When $\omega_1=1$ and $\omega_2$ is a pure imaginary number, these operators
are unitary with respect to the hermitian form (\ref{def:herm.form:y}).

%
%
%
%


\begin{thebibliography}{99}

\bibitem{gaudin:76}
Gaudin,~M.,
{\em Diagonalisation d'une classe d'hamiltoniens de spin,}
J. de Physique {\bf 37} (1976) 1087--1098.

\bibitem{gaudin:83}
Gaudin,~M.,
{\em La fonction d'onde de Bethe,} Paris: Masson (1983).

\bibitem{baxter:book}
Baxter,~R.~J.,
{\em Exactly Solved Models of Statistical Mechanics,}
NY: Academic Press (1982).

\bibitem{skl-tak:96}
Sklyanin,~E.\ K., Takebe,~T.,
Algebraic Bethe Ansatz for XYZ Gaudin model.
{\em Phys. Lett. A} {\bf 219},
(1996),
217--225.

\bibitem{b-ls-p:98}
Babujian, H., Lima-Santos, A., Poghossian, R. H.,
Knizhnik-Zamolodchikov-Bernard equations connected with the eight-vertex
model.
preprint {\tt solv-int/9804015}.

\bibitem{skl:87}
Sklyanin, E.~K.,
Separation of Variables in the Gaudin model,
{\em Zapiski Nauchnykh Seminarov LOMI} {\bf 164}
(1987),
151--169,
(in Russian);
{\em J. Sov. Math.} {\bf 47}
(1989),
2473--2488,
(English transl.).

\bibitem{skl:95}
Sklyanin, E.~K.,
Separation of variables. New trends.
in {\em Quantum field theory, integrable models and beyond
(Kyoto, 1994). 
Progr. Theoret. Phys. Suppl.} {\bf 118}
(1995), 
35--60.

\bibitem{fre:95}
Frenkel, E.,
Affine algebras, Langlands duality and Bethe ansatz. 
{\em {\rm XI}th International Congress of Mathematical Physics}
(Paris, 1994),
606--642,
Internat. Press, Cambridge, MA,
(1995).

\bibitem{e-f-r:98}
Enriquez, B., Feigin, B., Rubtsov, V.,
Separation of variables for Gaudin-Calogero systems.
{\em Compositio Math.} {\bf 110}
(1998),
1--16.

\bibitem{kur-tak:97}
Kuroki,~G., Takebe,~T.,
Twisted Wess-Zumino-Witten models on elliptic curves.
{\em Commun. Math. Phys.} {\bf 190}
(1997), 
1--56.

\bibitem{skl:86}
Sklyanin, E.~K.,
On Poisson structure of periodic classical $XYZ$-chain,
{\em Zapiski Nauchnykh Seminarov LOMI} {\bf 150}
(1986),
154--180,
(in Russian);
{\em J. Sov. Math.} {\bf 46}
(1989),
1664--1683,
(English transl.).

\bibitem{takh-fad:79}
Takhtajan,~L.~A. and Faddeev,~L.~D.,
The quantum method of the inverse problem and the Heisenberg XYZ Model,
{\em Uspekhi Mat. Nauk} {\bf 34:5}
(1979),
13--63,
(in Russian);
{\em Russian Math. Surveys} {\bf 34:5}
(1979),
11--68,
(English transl.).

\bibitem{whit-wat}
Whittaker, E.~T.~and Watson, G.~N.,
{\em A Course of Modern Analysis}, 4th ed.,
Cambridge University Press,
(1927).

\bibitem{mum}
Mumford,~D.,
Tata Lectures on Theta I.
Basel-Boston: Birkh\"auser,
1982

\bibitem{dag-sch:96}
D'Agnolo,~A., Schapira,~P.,
Radon-Penrose transform for $D$-modules.
{\em J. Funct. Anal.} {\bf 139}
(1996),
349--382.

\bibitem{dick}
Dickson, L. E.,
{\em Elementary theory of equations},
John Wiley \& Sons, Inc.,
(1914).

\bibitem{craw}
Crawford, L.,
On the solution of Lam\'e's equation
$d^2U/du^2 = U\{n(n+1) pu + B\}$ in finite terms when $2n$ is an odd
number,
{\em Quarterly J.~Pure and Appl.~Math.}, {\bf XXVIII}
(1895),
93--98.

\bibitem{skl:83}
Sklyanin, E.~K.,
Some Algebraic Structures Connected with the Yang-Baxter Equation.
Representations of Quantum Algebras,
{\em Funkts. analiz i ego Prilozh.} {\bf 17-4}
(1983),
34--48,
(in Russian);
{\em Funct. Anal. Appl.} {\bf 17}
(1984),
273--284,
(English transl.).


\end{thebibliography}
\end{document}